# Renormalons and Analytic Properties of the β Function

**I. M. Suslov***

*P.L. Kapitsa Institute of Physical Problems, Russian Academy of Sciences, Moscow, 117334 Russia*
Received April 19, 2004

**Abstract**—The presence or absence of renormalon singularities in the Borel plane is shown to be determined by the analytic properties of the Gell-Mann–Low function $\beta(g)$ and some other functions. A constructive criterion for the absence of singularities consists in the proper behavior of the $\beta$ function and its Borel image at infinity, $\beta(g) \propto g^\alpha$ and $B(z) \propto z^\alpha$ with $\alpha \leq 1$. This criterion is probably fulfilled for the $\varphi^4$ theory, quantum electrodynamics, and quantum chromodynamics, but is violated in the $O(n)$-symmetric sigma model with $n \longrightarrow \infty$.

(1) More than twenty years ago, Lipatov [1] suggested a method for calculating the high orders of the perturbation theory according to which these are determined by the saddle-point configurations (instantons) of the corresponding functional integrals. The method proved to be applicable to a wide range of problems [2, 3], but it was soon questioned in connection with the detection of factorially large contributions from the individual diagrams, renormalons [4]. In the opinion of 't Hooft [5], the latter are not contained in the instanton contribution. Formally, the asymptotics of the perturbation theory is determined by the singularity in the Borel plane closest to the coordinate origin. Whereas the presence of instanton singularities is beyond doubt, the existence of renormalon singularities has never been proven, which is recognized by the most active proponents of this trend [6]: such singularities can be easily obtained by the summation of individual sequences of diagrams, but it cannot be made sure that they are preserved when all diagrams are taken into account. Previously [7], we presented a proof for the absence of renormalon singularities in the $\varphi^4$ theory, which calls into question the idea of renormalons as a whole; however, there is no analogous proofs for other field theories. The analysis performed below clarifies the situation with renormalon singularities in an arbitrary field theory: in general, their presence or absence is determined by the analytic properties of the Gell-Mann–Low function and some other functions.

The simplest class of renormalon diagrams arises in quantum electrodynamics after distinguishing of an internal photon line in an arbitrary diagram (Fig. 1a) and insertion of a chain of electron loops into it (Fig. 1b). In the initial diagram, the integration $\int d^4k k^{-2n}$ over the range of large momenta ($n = 3, 4, \ldots$) corresponded to the separated photon line with momentum $k$. When $N$ electron loops are inserted into the photon line, the additional factor $\ln^N(k^2/m^2)$ ($m$ is the electron mass) emerges in the integrand; the integration yields a quantity on the order of $N!$. Arbitrary insertions into the photon line lead to the substitution of the running coupling constant $g(k^2)$ for the interaction constant $g_0$ (Fig. 1c) and give rise to the integral $\int d^4k k^{-2n} g(k^2)$. The summation of the chains of loops corresponds to using the single-loop approximation $\beta(g) = \beta_2 g^2$ for the Gell-Mann–Low function and yields the well-known result

$$g(k^2) = \frac{g_0}{1 - \beta_2 g_0 \ln(k^2/m^2)}. \quad (1)$$

After the integration over $k^2 \gtrsim m^2$, we obtain

$$\int d^4k k^{-2n} g(k^2) = g_0 \sum_N \int d^4k k^{-2n} \left(\beta_2 g_0 \ln \frac{k^2}{m^2}\right)^N$$
$$\sim g_0 \sum_N N! \left(\frac{\beta_2}{n-2}\right)^N g_0^N. \quad (2)$$

After the Borel summation, this yields renormalon sin-

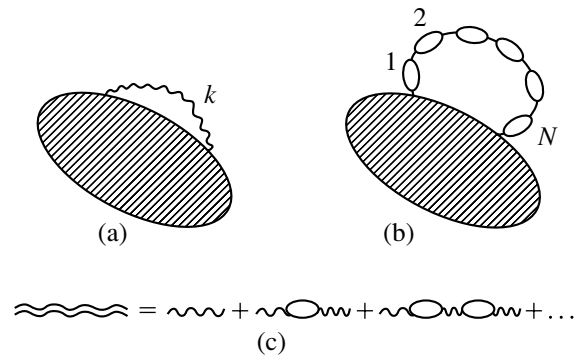

**Fig. 1.**

gularities at the points[1]

$$z_n = \frac{n-2}{\beta_2}, \quad n = 3, 4, 5, \ldots \quad (3)$$

in the Borel $z$ plane.

The analysis performed below is based on the fact that for a given β function, the summation of the entire class of diagrams obtained by all the possible insertions into the photon line presents no problem: it will suffice to solve the Gell-Mann–Low equation

$$\frac{dg}{d\ln k^2} = \beta(g) = \beta_2 g^2 + \beta_3 g^3 + \ldots \quad (4)$$

with the initial condition $g(k^2) = g_0$ at $k^2 = m^2$ and to analyze the expansion in terms of $g_0$ for an integral of type (2). The more complex classes of renormalon diagrams can be studied by using the general equation of a renormalization group in the Callan–Symanzik form.

(2) As an illustration, let us consider the model β function

$$\beta(g) = \frac{\beta_2 g^2}{1 + g^2}, \quad (5)$$

for which Eq. (4) can be easily solved:

$$g(k^2) = -\frac{1}{2}\left(\frac{1}{g_0} - g_0 - x\right) + \sqrt{\frac{1}{4}\left(\frac{1}{g_0} - g_0 - x\right)^2 + 1}, \quad (6)$$

where $x = \beta_2 \ln(k^2/m^2)$. The right-hand side is regular at $g_0 = 0$ and can be expanded into a power series of $g_0$. The structure of this series is

$$g = \sum_{N=1}^{\infty} A_N \left\{\frac{g_0}{r(x)}\right\}^N, \quad (7)$$

where $r(x)$ is the radius of convergence, and the coefficients $A_N$ depend on $N$ as a power law. The radius of convergence is determined by the distance to the singularity closest to the coordinate origin. The singularities $g_c$ in the right-hand side of Eq. (6) correspond to zeros of the radicand and are defined by the equation

$$g_c^2 + (x + 2i)g_c - 1 = 0 \quad (8)$$

and its complex conjugate. At large $x$, the minimum (in the absolute value) root is $g_c \approx 1/x$, and series (7) takes the form

$$g(k^2) = \sum_{N=1}^{\infty} A_N (g_0 x)^N = \sum_{N=1}^{\infty} A_N \left(\beta_2 \ln\frac{k^2}{m^2}\right)^N g_0^N. \quad (9)$$

Substituting it into integral (2) yields singularities at points (3) (at large $N$, the integral is determined by large $k$ that correspond to large $x$). Thus, reasoning by Parisi [8] that renormalon singularities can exist in internally consistent theories is confirmed by this example.

---
[1] Similar singularities with $n = 0, -1, -2, \ldots$ arise from the integration over the range of small momenta (infrared renormalons).

(3) The overall picture is determined by the fate of the Landau pole in the single-loop result (1). This pole can remain on the real axis, shift into the complex plane, or go to infinity. The right-hand side of Eq. (1) as a function of $g_0$ changes on the characteristic scale $(\ln(k^2/m^2))^{-1}$; this property does not change when the higher loops are taken into account, because result (1) is always valid at small $g_0$. If $g(k^2)$ as a function of $g_0$ has singularities in a finite part of the complex plane, then the characteristic scale of its change is naturally determined by the distance to the closest singularity, which thus proves to be of the order $(\ln(k^2/m^2))^{-1}$, generating a series of type (9) and renormalon singularities. However, this is not always the case: for example, the characteristic scale of the change for entire functions is determined by other factors, and the above conclusion ceases to be valid.

The general solution of the Gell-Mann–Low equation is

$$F(g) = F(g_0) + \ln\frac{k^2}{m^2}, \quad (10)$$

where

$$F(g) = \int \frac{dg}{\beta(g)}.$$

Taking into account the behavior of the function $F(g)$ at small $g$, we can write

$$F(g) = -\frac{1}{\beta_2 g} + f(g), \quad \text{where} \quad \lim_{g \to 0} gf(g) = 0, \quad (11)$$

and, formally resolving (10) for $g$, obtain

$$g(k^2) = F^{-1}\left\{-\frac{1}{\beta_2 g_0} + f(g_0) + \ln\frac{k^2}{m^2}\right\}. \quad (12)$$

If the function $z = F(g)$ is regular at $g_0$ and $F'(g_0) \neq 0$, then the inverse function $g = F^{-1}(z)$ that is also regular exists in some neighborhood of the point $g_0$. Therefore, the singularities of the function $F^{-1}(z)$ are $z_c = F(g_c)$, where all the possible $g_c$ are defined by the condition

$$F'(g_c) = 0 \text{ or } F'(g_c) \text{ does not exist.} \quad (13)$$

The singularities in variable $g_0$ in (12) are defined by the equation

$$z_c = -\frac{1}{\beta_2 g_0} + f(g_0) + \ln\frac{k^2}{m^2} \quad (14)$$

or

$$g_0 x - 1 = \beta_2 g_0 [z_c - f(g_0)]. \quad (15)$$

If $z_c$ is finite, then Eq. (15) at large $x$ has a root $g_0 \approx 1/x$ in the range of small $g_0$, where the right-hand side of

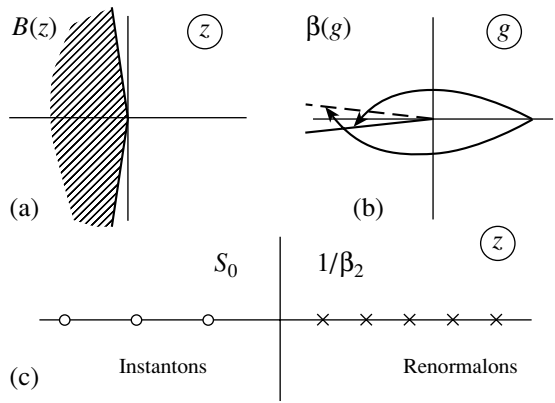

**Fig. 2.** It follows from the analyticity of the function $B(z)$ at $|\arg z| < \pi/2 + \delta$ (a) that $\beta(g)$ is analytic at $|\arg g| < \pi + \delta$ (b), i.e., on the entire physical sheet of the Riemann surface; (c) the picture of singularities in the Borel plane for the $\varphi^4$ theory and quantum electrodynamics suggested by 't Hooft ($S_0$ is the minimum instanton action, $\beta_2$ is the first vanishing expansion coefficient of the $\beta$ function).

Eq. (15) is insignificant in view of (11). Thus, there is a singularity at $g_c \approx 1/x$ that generates series (9) leading to the renormalon singularities (3). If, alternatively, $z_c = \infty$, then Eq. (14) has no solutions for $g_0 \sim 1/x$, and an expansion of type (9) is possible only with the coefficients $A_N$ that decrease faster than any exponential: the renormalon contribution is definitely much smaller than the instanton contribution, and no singularities emerge in the Borel plane. As for solutions with $g_0 \sim 1$ (which are possible due to the singularities of the function $f(g_0)$), they are unrelated to the renormalon mechanism: their contribution is determined by a series in which $g_0^N$ is not accompanied by a factor of the type $(\ln(k^2/m^2))^N$. To summarize, we have reached the following conclusion. Renormalon singularities take place if there exists at least one point $g_c$ (including $g_c = \infty$) for which condition (13) is satisfied and $z_c = F(g_c) < \infty$; otherwise, no renormalon singularities exist.

It remains to reformulate the results in terms of the $\beta$ function itself. First, note that a regular root of the form

$$\beta(g) \sim (g - g_c)^n, \quad n = 1, 2, 3, \ldots$$

does not lead to renormalons: in this case, the derivative $F'(g_c)$ does not exist, but $F(g_c) = \infty$; in particular, this is true for the root at $g = 0$. A power-law behavior at infinity, $\beta(g) \propto g^\alpha$, generates renormalons only at $\alpha > 1$ (which is identical to the existence condition for the Landau pole for a nonalternating function $\beta(g)$). All of the other possibilities for the satisfaction of condition (13) are related to the singularities of the function $\beta(g)$ at finite points $g_c$: for renormalons to exist, these must be strong enough for the function $1/\beta(g)$ to be integrable at $g_c$ (e.g., $\beta(g) \propto (g - g_c)^\gamma$ with $\gamma < 1$). A sufficient condition for the absence of renormalons is the regularity of the function $\beta(g)$ at finite $g$ and its power-law behavior, $\beta(g) \propto g^\alpha$ with $\alpha \leq 1$, at infinity; in fact, weak singularities of the type $\beta(g) \propto (g - g_c)^\gamma$ with $\gamma > 1$ are admissible at finite $g$.

(4) If all of the singularities in the Borel plane are assumed to be of instanton or renormalon origin,[2] then a constructive criterion for the absence of renormalon singularities can be formulated.

The perturbation series for the $\beta$ function is factorially divergent [1–3], and there is a cut in the complex $g$ plane that emerges from the coordinate origin to infinity. Therefore, $g = 0$ is the branching point, as is generally $g = \infty$. The function $\beta(g)$ is represented by the Borel integral

$$\beta(g) = \int_0^\infty dz\, e^{-z} B(gz) = g^{-1} \int_0^\infty dz\, e^{-z/g} B(z), \qquad (16)$$

where $B(z)$ is the Borel image of the function $\beta(g)$. Let us assume that it has a power-law behavior at infinity, $B(z) \propto z^\alpha$ (then $\beta(g) \propto g^\alpha$), and is regular for $|\arg z| < \pi/2 + \delta$, $\delta > 0$ (Fig. 2a). Directing the contour of integration along the ray $z = |z| e^{i\phi_0}$, we can easily verify that integral (17) converges for $g = |z| e^{i\phi_0}$ with $|\phi - \phi_0| < \pi/2$. Since the contour can turn through angles $|\phi_0| < \pi/2 + \delta$, the function $\beta(g)$ is regular for $|\arg g| < \pi + \delta$ (Fig. 2b), implying that there are no singularities at finite points on the physical sheet of the Riemann surface. In this case, the behavior of the $\beta$ function at infinity ($\beta(g) \propto g^\alpha$ with $\alpha \leq 1$) yields the condition for the absence of renormalon singularities.

The derived criterion can be constructively used as follows. Consider the $\varphi^4$ theory or quantum electrodynamics; in this case, there are instanton singularities on the negative semiaxis and, possibly [5], renormalon singularities on the positive semiaxis (Fig. 2c). Let us assume that there are no renormalon singularities. In this case, (i) the regularity condition for the function $\beta(g)$ at finite $g$ (Figs. 2a and 2b) is satisfied; (ii) the asymptotics of the expansion coefficients is determined by the nearest instanton singularity and can be found by Lipatov's method; (iii) the Borel integral is well defined, and the perturbation series for the function $\beta(g)$ admits an unambiguous summation, which allows its behavior at infinity to be determined. If the $\beta$ function increases faster than $g^\alpha$ with $\alpha > 1$, then the initial assumption is invalid, and the existence of renormalon singularities has been proven by contra-diction. If, alternatively, $\beta(g) \propto g^\alpha$ with $\alpha < 1$, then the assumption about the absence of renormalon singularities is self-consistent.

---

[2] This assumption has not been rigorously proven, but nobody has proposed a viable alternative to it. It can be justified by the fact that all of the singularities in the Borel plane for finite-dimensional integrals are related to the extrema of the action (in this case, the reasoning of 't Hooft in [5] is necessary and sufficient), while the renormalon singularities are explicitly related with the limiting transition to an infinite number of integrations.

The outlined program for the above theories was carried out previously [9, 10] by interpolating Lipatov's asymptotics with known values of the first expansion coefficients and yielded $\alpha = 0.96 \pm 0.01$ for the $\varphi^4$ theory [9] and $\alpha = 1.0 \pm 0.1$ for quantum electrodynamics [10]. Thus (within the uncertainty of the results), the self-consistent exclusion of renormalon singularities proves to be possible. Moreover, a comparison with existing analytic results is indicative of the exact equality $\alpha = 1$ in both cases [9, 10]. In any case, the β functions in these theories are nonalternating,[3] and the condition for the absence of renormalon singularities in them is identical to the condition for their internal consistency.

For quantum chromodynamics, $\alpha = -12 \pm 3$ [11], and the instanton singularities lie on the positive semiaxis. The assumption about the absence of renormalon singularities is self-consistent for the sheet of the Riemann surface obtained by the analytic continuation from negative $g$ (the sign of the argument of B(z) changes as the sign of $g$ changes, and the singularities pass to the negative semiaxis); this is enough to justify the procedure for determining[4] the index $\alpha$ used in [11]. The Borel integral at positive $g$ must be properly interpreted to establish a connection with the physical sheet (its principal-value interpretation is not always correct [12]).

The only field theory in which the existence of renormalon singularities is deemed to have been firmly established is the $O(n)$-symmetric sigma model in the limit $n \longrightarrow \infty$ [6]. In this case, the single-loop β function is exact and $\beta(g) \propto g^2$ for $g \longrightarrow \infty$; consequently, $\alpha = 2$ and the self-consistent exclusion of renormalons proves to be impossible. However, this theory is internally inconsistent in the four-dimensional case.

Curiously, according to the formulated criterion, the truncation of the series for the β function at any finite number of terms immediately creates renormalon singularities. This shows that the problem of renormalons cannot be solved in terms of the loop expansion [13].

Note that the possibility of the existence of renormalon singularities makes the functional integrals ill-defined. The classical definition of the functional integral via the perturbation theory is defective, because the expansion in terms of the coupling constant is divergent: its constructive summation requires knowing the analytic properties in the Borel plane that are uncertain until it is established whether the renormalon singularities exist. One can also doubt that the definition of the functional integral as a multidimensional integral on a lattice is correct; the lattice theory can differ fundamentally from the continuum theory, because the renormalon contributions are determined by the range of arbitrarily large momenta. An impasse is reached: the solution of the problem of renormalons requires studying the functional integrals, while the latter are ill-defined because the problem of renormalons is unsolved. The proposed scheme for the self-consistent exclusion of renormalon singularities is probably the only possible way out of the situation. In this case, the continuum theory, by definition, is understood to be the limit of the lattice theories.

(5) In general, a class of renormalon diagrams is defined by the condition that new vertices are inserted into the same element (a line or a vertex) of the initial skeleton diagram. This definition allows the existence conditions for the main renormalon contribution to be analyzed: if new vertices are inserted with an equal probability into $m$ different elements, then the corresponding contribution is on the order of $[(N/m)!]^m \sim N! m^{-N}$ and contains the redundant smallness $m^{-N}$.[5] For electrodynamics, integral (2) considered above corresponds to the summation of the class of diagrams obtained by all the possible insertions into the same photon line. A similar integral for the $\varphi^4$ theory corresponds to all the possible loop insertions to the same 4-vertex and the domain of integration is considered, in which all momenta of the considered vertex are of the same order of magnitude. In general, the renormalon integral is

$$\int d^4 k \, k^{-2n} \Gamma(g_0, k), \quad (17)$$

where $\Gamma(g_0, k)$ is the vertex with $M$ external lines from which $M'$ lines carry a large momentum of the order $k$. The dependence on $k$ is defined by the Callan–Symanzik equation

$$\left[-\frac{\partial}{\partial \ln k^2} + \beta(g_0)\frac{\partial}{\partial \ln g_0} + \gamma(g_0)\right]\Gamma(g_0, k) = 0, \quad (18)$$

where $\gamma(g_0)$ depends on $M$ and $M'$. The general solution of Eq. (18) is

$$\Gamma(g_0, k) = \exp F_1(g_0) \Phi\left(F(g_0) + \ln\frac{k^2}{m^2}\right),$$

$$F_1(g) = -\int dg \frac{\gamma(g)}{\beta(g)},$$

$$(19)$$

where $\Phi(z)$ is an arbitrary function. If $z_c$ is a singularity of the function $\Phi(z)$, then the singularities in variable $g_0$

---

[3] For the $\varphi^4$ theory, we have in mind the four-dimensional case, in which the problem of renormalons is of current interest.

[4] Note that the asymptotics $\beta(g) \propto g^\alpha$ does not guarantee a power-law behavior of the Borel image, $B(z) \propto z^\alpha$, in all directions in the complex plane (e.g., $B(z) = \beta_2(1 - \cos z)$ for the model β function (5)). In this respect we should emphasize that the index $\alpha$ in [9–11] was determined directly from the asymptotics of the Borel image and the assumption about its power-law behavior was subjected to the special tests.

[5] In fact, in this case, there are $m$ independent integrations of type (17) for each of which the condition for the absence of renormalon contributions is identical to that established below.

are defined by Eq. (14). The function $\Phi(z)$ can be expressed in terms of $R(g_0) \equiv \Gamma(g_0, m)$,

$$\Phi(z) = R(F^{-1}(z)) \exp\{-F_1(F^{-1}(z))\}, \qquad (20)$$

and the singularities of the function $F^{-1}(z)$ are those of the function $\Phi(z)$. Therefore, the condition for the existence of renormalons found above is also sufficient in the general case. Additional possibilities for their emergence are associated with the singularities of the functions $F_1(g)$ and $R(g)$. If one of them is singular at $g_c$, then $z_c = F(g_c)$ is a singularity of the function $\Phi(z)$. The functions $F_1(g)$ and $R(g)$ are represented by Borel integrals of type (16) and have $g = 0$ and $\infty$ as the branching points. However, this does not lead to singularities of the function $\Phi(z)$ at finite points, because

$$F(0) = \infty, \quad F(\infty) = \infty \quad (\text{for } \alpha \le 1).$$

At the same time, the singularities at finite $g$ can be self-consistently excluded for the functions $F_1(g)$ and $R(g)$, as it was done above for the $\beta$ function. As a result, the behavior of the function $\beta(g)$ at infinity also determines the presence or absence of renormalons in the general case.

(6) It is clear from the above discussion that using information only from the renormalization group, we can establish the necessary and sufficient conditions for the existence of renormalons, but cannot come to any definite conclusions. Let us compare this with Parisi's renormalization-group analysis [8] that underlies all of the recent studies devoted to renormalons [6]. If, following [8], the momentum dependence of the Borel images is assumed to differ from the single-loop result only by a slowly changing factor, then this ansatz formally satisfies the equations if we expand the slowly changing function in terms of gradients and restrict the analysis to the local approximation. However, to study the stability of the solution, we should continue the expansion in gradients and obtain a diffusion-type equation. The solution is stable if the corresponding diffusion coefficient is positive, which, in general, is not the case. The extent, to which Parisi's solution is destroyed, is determined by the rather subtle properties of the $\beta$ function, which correlates with the above assertions.

In conclusion, let us discuss the subtle point in the proof for the $\varphi^4$ theory that was inadequately covered in [7]. Any quantity defined by the perturbation series is a function of the bare charge $g_B$ and the cutoff parameter $\Lambda$. Passing to the renormalized charge $g$ gives rise to the function $F(g, \Lambda)$ that contains the residual dependence on $\Lambda$, but has the finite limit in view of the renormalizability:

$$\lim_{\Lambda \to \infty} F(g, \Lambda) = F(g). \qquad (21)$$

A similar property is expected for the corresponding Borel images:

$$\lim_{\Lambda \to \infty} B(z, \Lambda) = B(z). \qquad (22)$$

The analyticity of the function $B(z, \Lambda)$ at finite $\Lambda$ in the complex $z$ plane with a cut from the first instanton singularity to infinity was rigorously proven previously [7]. The function $B(z)$ is analytic in the same domain under the condition of uniform convergence in (22) (the Weierstrass theorem [14]); the latter takes place if the function $B(z, \Lambda)$ is bounded (the compactness principle for regular functions [15]). Therefore, the finiteness of the limit in (22) is enough to prove[6] the regularity of the function $B(z)$.

Unfortunately, the finiteness of the limits in (21) and (22) has been rigorously proven only in terms of the perturbation theory, i.e., not for the functions $F(g, \Lambda)$ and $B(z, \Lambda)$ themselves, but for the coefficients of their expansion in $g$ and $z$. The proof in [7] suggests that the limits are finite at the level of functions, and, in this sense, it is incomplete. However, the finiteness of the limits in (21) and (22) is required for the existence of true renormalizability and should be considered as a necessary physical condition for it. The latter is closely related to the necessity of redefining the functional integrals noted above.

## ACKNOWLEDGMENTS

I am grateful to L.N. Lipatov for numerous discussions on the problem of renormalons during which the idea of this work arose. This study was supported by the Russian Foundation for Basic Research (project no. 03-02-17519).

## REFERENCES

1. L. N. Lipatov, Zh. Éksp. Teor. Fiz. **72**, 411 (1977) [Sov. Phys. JETP **45**, 216 (1977)].
2. *Large-Order Behavior of Perturbation Theory*, Ed. by J. C. Le Guillou and J. Zinn-Justin (North-Holland, Amsterdam, 1990).

---

[6] If the function $B(z)$ has a singularity at $z_0$ that is the point of regularity for $B(z, \Lambda)$, then the function $B(z, \Lambda)$ is unbounded for $\Lambda \longrightarrow \infty$ on any contour that encloses the point $z_0$. This unboundedness is global in nature and is unrelated to the possible divergence of the function $B(z)$ at $z = z_0$. In fact, if the Borel image of $B(z)$ becomes infinite at isolated points in the complex plane, then it is easy to ensure its boundedness for the most interesting case of the power-law singularities. Let, by contradiction, $B(z) \propto (z - z_0)^{-\gamma}$ in the neighborhood of $z_0$ that is the point of regularity for $B(z, \Lambda)$. Let us turn to a more general definition of the Borel image in which the coefficients of the initial series are divided not by $N!$, but by $\Gamma(N + b_0)$; the index $\gamma$ can then be made negative by increasing $b_0$ [7, Section 3.1]. Thus, the function $B(z)$ becomes bounded near $z_0$, and no singularity can exist at $z = z_0$. The regularity at arbitrary $b_0$ follows from the regularity of $B(z)$ at large $b_0$ [7, Section 3.1].

3. E. B. Bogomolny, V. A. Fateyev, and L. N. Lipatov, in *Soviet Science Reviews, Section A: Physics Reviews,* Ed. by I. M. Khalatnikov (Harwood Academic, New York, 1980), Vol. 2, p. 247.
4. B. Lautrup, Phys. Lett. B **69**, 109 (1977).
5. G. 't Hooft, in *The Whys of Subnuclear Physics, Erice, 1977*, Ed. by A. Zichichi (Plenum, New York, 1979).
6. M. Beneke, Phys. Rep. **317**, 1 (1999).
7. I. M. Suslov, Zh. Éksp. Teor. Fiz. **116**, 369 (1999) [JETP **89**, 197 (1999)].
8. G. Parisi, Phys. Rep. **49**, 215 (1979).
9. I. M. Suslov, Zh. Éksp. Teor. Fiz. **120**, 5 (2001) [JETP **93**, 1 (2001)].
10. I. M. Suslov, Pis'ma Zh. Éksp. Teor. Fiz. **74**, 211 (2001) [JETP Lett. **74**, 191 (2001)].
11. I. M. Suslov, Pis'ma Zh. Éksp. Teor. Fiz. **76**, 387 (2002) [JETP Lett. **76**, 327 (2002)].
12. R. Seznec and J. Zinn-Justin, J. Math. Phys. **20**, 398 (1979).
13. F. David, Nucl. Phys. B **209**, 433 (1982); Nucl. Phys. B **234**, 237 (1984); Nucl. Phys. B **263**, 637 (1986).
14. G. A. Korn and T. M. Korn, *Mathematical Handbook for Scientists and Engineers*, 2nd ed. (McGraw-Hill, New York, 1968; Nauka, Moscow, 1977).
15. M. A. Evgrafov, *Analytic Functions,* 2nd ed. (Nauka, Moscow, 1968; Saunders, Philadelphia, Pa., 1966).

*Translated by V. Astakhov*